\documentclass[fleqn,usenatbib]{mnras}
\usepackage{newtxtext,newtxmath}
\usepackage[T1]{fontenc}
\usepackage{ae,aecompl}

\usepackage{graphicx}	
\usepackage{balance}
\usepackage{multirow}
\usepackage{siunitx}

\DeclareSIUnit\parsec{pc}

\usepackage[usenames,dvipsnames]{xcolor}
\hypersetup{
    colorlinks = true,
    citecolor = {MidnightBlue},
    linkcolor = {BrickRed},
    urlcolor = {BrickRed}
}

\newcommand{\be}{\begin{equation}}
\newcommand{\ee}{\end{equation}}
\newcommand{\bea}{\begin{eqnarray}}
\newcommand{\eea}{\end{eqnarray}}

\usepackage{geometry}
\title[Doppler magnification in an ensemble of relativistic simulations]{Observing relativistic features in large-scale structure surveys -- II: Doppler magnification in an ensemble of relativistic simulations}

\author[L.~Coates et al.]{Louis Coates,$^{1}$\thanks{l.j.c.coates@qmul.ac.uk}
Julian Adamek,$^{1,2}$ 
Philip Bull,$^{1,3}$ 
Caroline Guandalin,$^{4}$ 
Chris Clarkson$^{1,3}$ 
\\
$^{1}$Astronomy Unit, Queen Mary University of London, Mile End Road, London, E1 4NS, UK\\
$^{2}$Institute for Computational Science, University of Zurich, Winterthurerstrasse 190, 8057 Zurich, Switzerland\\
$^{3}$Department of Physics \& Astronomy, University of the Western Cape, Cape Town 7535, South Africa\\
$^{4}$Departamento de F\'isica Matem\'atica, Instituto de F\'isica, Universidade de S\~ao Paulo, Rua do Mat\~ao 1371, CEP 05508-090, S\~ao Paulo, Brazil
}

\date{Accepted XXX. Received YYY; in original form ZZZ}

\pubyear{2020}

\begin{document}
\label{firstpage}
\maketitle

\begin{abstract}
The standard cosmological model is inherently relativistic, and yet a wide range of cosmological observations can be predicted accurately from essentially Newtonian theory. This is not the case on `ultra-large' distance scales, around the cosmic horizon size, however, where relativistic effects can no longer be neglected. In this paper, we present a novel suite of 53 fully relativistic simulations generated using the \textit{gevolution} code, each covering the full sky out to 
$z \approx 0.85$,
and approximately 1930 square degrees out to 
$z \approx 3.55$. These include a relativistic treatment of massive neutrinos, as well as the gravitational potential that can be used to exactly calculate observables on the past light cone. The simulations are divided into two sets, the first being a set of 39 simulations of the same fiducial cosmology (based on the Euclid Flagship 2 cosmology) with different realisations of the initial conditions, and the second which fixes the initial conditions, but varies each of seven cosmological parameters in turn. Taken together, these simulations allow us to perform statistical studies and calculate derivatives of any relativistic observable with respect to cosmological parameters. As an example application, we compute the cross-correlation between the Doppler magnification term in the convergence, $\kappa_v$, and the CDM+baryon density contrast, $\delta_{\rm cb}$, which arises only in a (special) relativistic treatment. We are able to accurately recover this term as predicted by relativistic perturbation theory, and study its sample variance and derivatives with respect to cosmological parameters.
\end{abstract}



\section{Introduction}

The large-scale distribution of matter in our Universe will be studied at ever greater detail with upcoming astronomical surveys such as LSST \citep{Abell:2009aa}, Euclid \citep{Laureijs:2011gra}, DESI \citep{Levi:2013gra}, and the Roman Telescope \citep{Spergel:2015sza}. The complicated astrophysical feedback processes that plague the dynamics of galaxies and clusters become subdominant at cosmological distance scales, making large-scale structure an ideal laboratory for probing gravity, which is effectively the only important force at those scales. This is usually done by extracting summary statistics such as power spectra, bispectra etc.\ \citep{Leclercq:2014jda, Bonvin:2011bg, Yoo:2009au}. As the volume of surveys increases, these will be measured more accurately and for an increasing number of modes. As a result, previously unconstrained small effects can be detected with high significance.

Cosmological $N$-body simulations provide a powerful and versatile means to predict these summary statistics given a cosmological model. These simulations commonly use Newtonian theory \citep{Springel:2005mi,Teyssier:2001cp} which is sufficient for many purposes, in particular in the context of the $\Lambda$CDM concordance model. At extremely large distance scales, the interpretation of such simulations becomes subtle, however \citep{Rigopoulos:2013nda,Green:2011wc,Chisari:2011iq}. One way of maintaining consistency with general relativity at leading order is by using so-called Newtonian motion gauges \citep{Fidler:2017pnb}, but the full machinery for analysing simulations in this context still needs to be developed.

Alternatively, general relativity can be implemented in the simulations explicitly. Employing techniques from numerical relativity, this has been explored e.g.\ in \citet{Giblin:2016mjp} and \citet{Macpherson:2016ict}. The main drawback of this formulation is the requirement to keep track of the wave-like solutions of the gravitational field, which needs extremely fine time resolution and thus leads to practical limitations. In the weak-field regime relevant for cosmology, however, one can easily perform a scalar-vector-tensor decomposition of the gravitational field in order to isolate these wave-like components. They can then be treated with fast approximate methods that completely remove the limitation on the time stepping. Such an approach is implemented in the weak-field relativistic $N$-body code \textit{gevolution} \citep{Adamek:2016zes,Adamek:2015eda} that we employ in this work. Relativistic effects that appear at extremely large distance scales are naturally included in our numerical simulations. Physical quantities, whenever they are gauge-dependent, are computed in the Poisson gauge, which is widely employed in practical calculations. Furthermore, using ray tracing we can compute physical observables exactly and directly.

While the final observables (i.e.\ the observed summary statistics) in actual surveys contain the fully aggregated information of how matter is distributed \textit{and} observed on our past light cone, in the weak-field regime it can often be useful to study different effects separately. In the literature, one often finds that any effects that are not explained by weak lensing or specifically the Kaiser-type redshift-space distortions \citep{Kaiser:1987qv} are called \textit{relativistic} effects, even though confusingly some of them would still be captured in Newtonian simulations, and even though weak lensing and redshift-space distortions are both arguably relativistic in nature as well. Some examples include corrections in the two-point correlation function, which can be as large as 10\% \citep{Lorenz:2017iez, Tansella:2017rpi, Bertacca:2012tp, Bonvin:2014owa, Beutler:2020evf, Raccanelli:2013gja,Yoo:2013tc}, or in bispectra where the signal-to-noise ratio for the relativistic part is $\sim$10 for a survey like Euclid \citep{Maartens:2019yhx, Jolicoeur:2020eup, deWeerd:2019cae, Clarkson:2018dwn, Jolicoeur:2017eyi, Umeh:2016nuh, Bertacca:2017dzm}. Such corrections are also relevant for the study of non-Gaussianity and bias \citep{Bruni:2011ta,Umeh:2019qyd,Wang:2020ibf,Alonso:2015uua,Camera:2014sba,Fonseca:2015laa}. The benefit of inherently relativistic simulations is that {\it all} such effects are transparently included, and so there is no ambiguity in the predictions for observable quantities.

In this paper, we present the UNITY simulations, a set of 53 fully relativistic $N$-body simulations for which we have retained \textasciitilde 35\,TB of data in the form of \textsc{HEALPix} maps of different fields as a function of comoving distance from a chosen observation point, and many power spectra etc. derived from these fields. A large portion of our simulations are run using the same fiducial cosmology, but using a varying random seed to generate the initial conditions, so as to give us multiple random realisations of the same underlying cosmology. The rest of the simulation suite contains pairs of simulations with the same random initial conditions, but each of the cosmological parameters varied by some percentage around the fiducial value. This allows us to compute numerical derivatives to determine the effect of each parameter on various summary statistics and observables. A full ray-tracing procedure can be applied to the stored data in order to produce light cones, or more selective treatments can be used to study particular fields and observables in isolation. The data are available on request\footnote{\href{http://philbull.com/unity}{http://philbull.com/unity}}.

The layout of the paper is as follows. In Sect.~\ref{sec:Simulations}, we describe the simulations in more detail, including the data products that are available. 
We then show some 
examples where the simulation data are used to extract a relativistic signal
in Sect.~\ref{sec:Results}, and then finally we conclude in Sect.~\ref{sec:Conclusions}.

\section{Simulations}\label{sec:Simulations}

\begin{table}
    \centering
    \caption{Details of the UNITY simulations. The first row is based on the fiducial cosmology, with multiple different realisations. This gives the ability to study the covariance. The second row is also run using the fiducial cosmology, but \textit{gevolution} is run in the $N$-body gauge. These can be used to compare and determine gauge effects in our simulation. Finally we list a series of pairs of simulations where we vary one of the parameters around the fiducial value. This allows us to calculate finite differences on different statistics.}
    \begin{tabular}{c c r l}
     {\bf No. sims} & {\bf Varied parameter} & {\bf Values}  & {\bf Gauge} \\ 
     \hline
     34 & Initial conditions          &     (random seed)       & Poisson \\
     5  & Initial conditions          &     (random seed)       & $N$-Body \\
     2  & $n_s$ & $0.69 \pm ~5\%$             & Poisson \\
     2  & $A_s$ & $2.1\times10^{-9}\pm ~5\%$  & Poisson \\
     2  & $h$ & $0.67\pm 5\%$                & Poisson \\
     2  & $\omega_b $ & $ 0.021996\pm 10\%$    & Poisson \\
     2  & $\omega_{\rm cdm} $ & $ 0.121203\pm 5\%$ & Poisson \\
     2  & $M_\nu $ & $ (0.1,0.2)$ eV            & Poisson \\
     2  & $w_0 $ & $-0.9$                     & Poisson \\
     \hline
    \end{tabular}
    \label{tab:simulations}
\end{table}

For our simulations, we use the $N$-body code \textit{gevolution}.
The code is described in full in \citet{Adamek:2016zes}, but we will give a brief overview of its workings here. \textit{gevolution} employs the Friedmann-Lema\^{i}tre-Robertson-Walker (FLRW) metric with perturbations in Poisson gauge and 
a weak-field setting where all gravitational fields ($\phi,\psi, B_i,$ and $h_{ij}$) are small. These metric quantities are stored on a Cartesian grid and are evolved 
together with the $N$-body ensemble that describes the cold dark matter (CDM) in phase space. The joint evolution is therefore computed using a particle-mesh approach, keeping a fixed and uniform resolution on the mesh.
The metric quantities can be saved on spatial hypersurfaces at any redshift alongside a full particle snapshot. The newest version of \textit{gevolution}\footnote{\href{https://github.com/gevolution-code/gevolution-1.2}{https://github.com/gevolution-code/gevolution-1.2}} also allows the data to be saved on a series of light cones given a specific (or several) observers. In this case, the metric quantities are saved on a series of \textsc{HEALPix}\footnote{\href{http://healpix.sourceforge.net}{http://healpix.sourceforge.net}} \citep{2005ApJ...622..759G, Zonca2019} maps in an approach that is better adapted to the geometry of the problem.

\subsection{The UNITY suite of simulations}\label{sec:UnitySuite}

Our simulations have a box volume of$ (\SI{4032}{\mega \parsec / \textit{h}})^3$, where all of the metric quantities are calculated on a Cartesian grid with $2304^3$ grid points. This means that we have spatial and mass resolutions of \SI{1.75}{\mega \parsec / \textit{h}} and $\approx \SI{4.6e11}{M_\odot}/h$, respectively. We use $2304^3$ particles to sample cold dark matter and baryons, while the massive neutrino species are treated with a grid-based approach similar to the one described in \citet{Brandbyge:2008js} that uses the linear transfer functions. For the case where we explore a non-standard equation of state parameter of dark energy, $w_0 = -0.9$ instead of $w_0 = -1$, we use a very similar approach to account for the perturbations of the dark energy fluid, see \citet{Dakin:2019vnj} and, in particular, \citet{Hassani:2019lmy} for details. As these perturbations are often neglected in Newtonian simulations of evolving dark energy, we also run a simulation for comparison where the dark energy is perfectly homogeneous (in Poisson gauge).

\begin{figure}
    \centering
    \includegraphics[width=0.8\columnwidth]{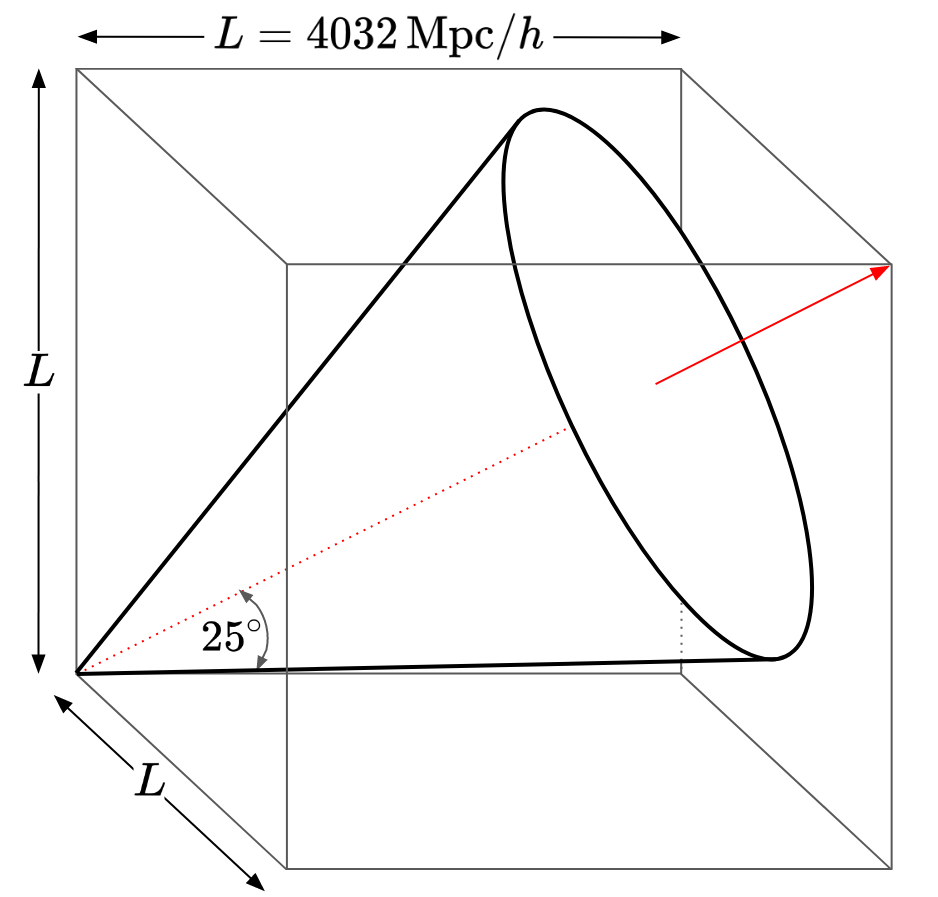}
    \caption{Schematic representation of the pencil-beam light cone construction. The observer is located at the vertex of the cone, with the line-of-sight along the diagonal of the simulation box (depicted by the red-dotted line and arrow). The opening half-angle of the pencil-beam light cone is 25 degrees.} 
    \label{fig:lc_construction}
\end{figure}

Each simulation has two light cones with the same observer at the corner of the box. The first light cone extends out to a distance of $\SI{2015}{\mega \parsec / \textit{h}}$ and covers the full sky\footnote{To build the full-sky case we use periodic boundary conditions, with each eighth of the sphere then}occupying a corner of the box., whereas the second one, illustrated in Fig.\ \ref{fig:lc_construction}, extends further out to a distance of $\SI{4690}{\mega \parsec / \textit{h}}$ with a disk-shaped survey area of approximately 1932 square degrees.The pointing at the centre of the disk is towards the opposite corner of the box to allow us the maximum distance without repeating data, which could add spurious correlations (especially on large scales). 

\begin{figure*}
    \centering
    \includegraphics[scale=0.48]{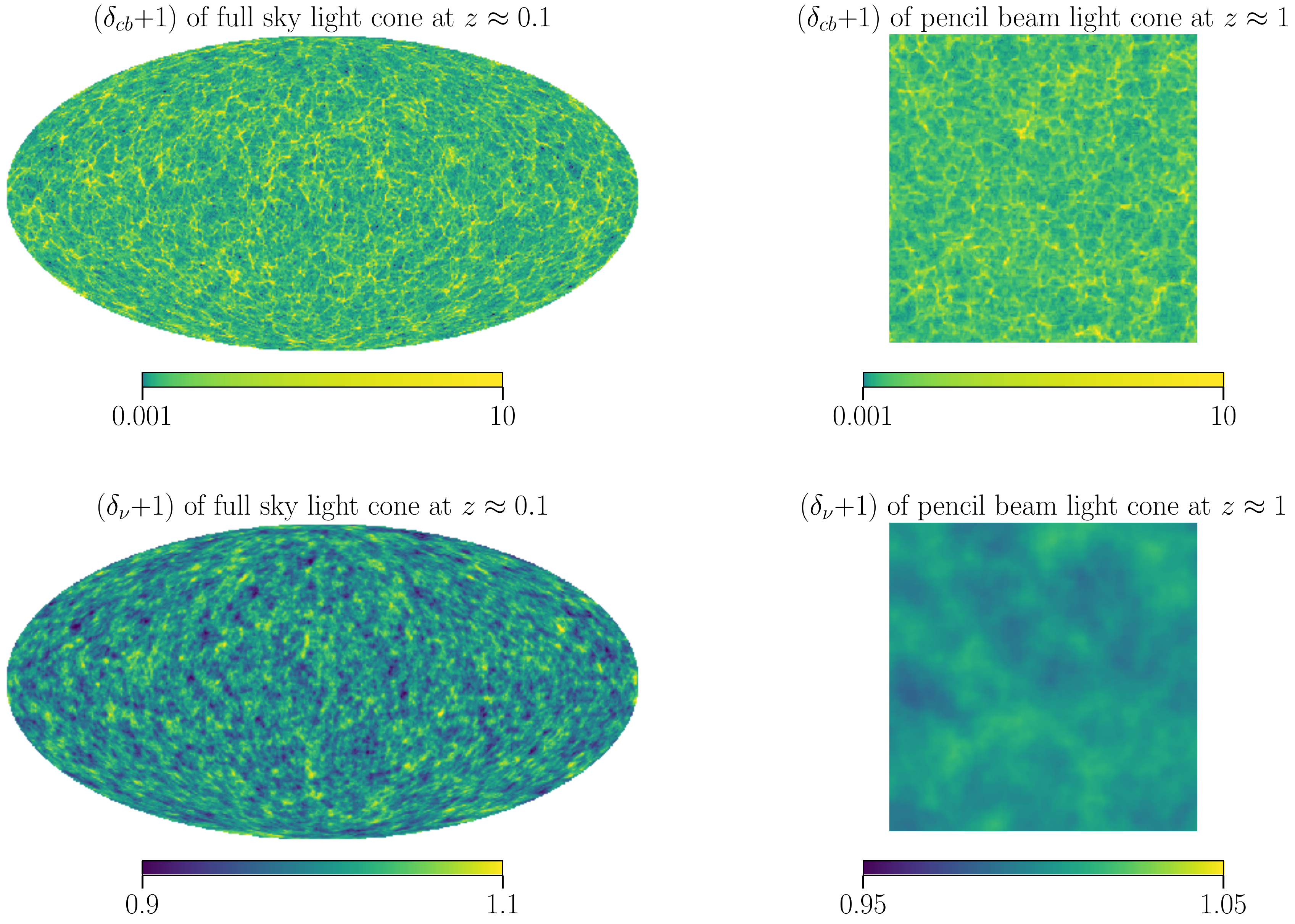}
    \caption{\textsc{HEALPix} maps for both the CDM+baryon density field and the neutrino density fields. We show these for the full-sky light cone at a redshift of $z\approx0.1$, and also for the pencil-beam light cone at a redshift of $z\approx1$. For the pencil beam we plot a $10^\circ \times 10^\circ$ section.}
    \label{fig:HealpixMaps}
\end{figure*}

The UNITY simulations consist of a total of 53 simulations, the parameters of which are summarised in Table \ref{tab:simulations}. Of these, 39 use the same fiducial cosmology of $n_s=0.96$, $A_s=\SI{2.1e-9}{}$, $h=0.67$, $\omega_b=0.021996$, $\omega_{cdm}=0.121203$, $M_\nu=0.06$ eV and $w_0=-1$. These parameters match the fiducial cosmology of the Euclid Flagship 2 simulations \citep{Knabenhans:2018cng}. For each of these simulations we vary the random seed, which gives us a different realisation of initial conditions for the same cosmology.

For the remaining simulations we vary several of the parameters individually, in turn, adding or subtracting a small amount from the fiducial value in such a way that we can form finite difference derivatives with respect to the parameters at a later stage,
\begin{equation}
    \frac{\partial \boldsymbol{S}}{\partial \theta} \simeq \frac{\boldsymbol{S}(\theta+\Delta \theta)-\boldsymbol{S}(\theta-\Delta \theta)}{2 \Delta \theta},
    \label{eq:finite_diff}
\end{equation}
where $\boldsymbol{S}$ is an observable quantity of interest and $\theta$ is the respective cosmological parameter. We varied each parameter by 5\%, as if $\Delta\theta$ is too small, the results for the finite difference will be dominated by numerical noise, since the difference between the observables will in many cases be minimal. For $\omega_b$ and $w_0$ we instead use a larger variation of 10\%. For the simulations where $M_{\nu}$ is varied we take a slightly different approach, since the neutrino mass scale is less well constrained and we would therefore like to allow for larger excursions. We considered cases with a total neutrino mass of 0.1 eV and 0.2 eV (while ensuring the same squared-mass differences between the three mass eigenstates). Note that we did not keep $\Omega_m$ fixed in these cases. 


\subsection{Data products}
\label{sec:DataProducts}

We retain several different types of data product from the simulations, giving us enough flexibility to calculate a wide range of observable quantities, but without needing to retain the full particle catalogue and metric perturbations on a grid for many snapshots, which would result in a very large data volume.

\paragraph*{\textsc{HEALPix} maps}---
Any calculations that we wish to do on the light cone are simplified by saving the data in spherical shells about a pre-specified observing location, instead of on a Cartesian grid for many snapshots in time. We use the HEALPix pixelisation to store the data for each field and for each shell. These maps are saved at a spatial resolution in radial slices of comoving width \SI{1.75}{\mega \parsec / \textit{h}} and cover both the full-sky and the pencil-beam light cone. The $N_{\rm side}$ of the maps varies as a function of the distance from the observer, always maintaining a sampling of the fields close to the resolution of the simulation and reaching a maximum of $N_{\rm side} = 2048$.

The quantities saved on these maps are the scalar potential $\phi$, the peculiar velocity field (coarse-grained at the scale of the pixel) projected onto the line of sight, the CDM+baryon ($\rm cb$) density, and the neutrino ($\nu$) density. These quantities are calculated, as usual in \textit{gevolution}, on a Cartesian grid, but are interpolated onto an appropriate set of \textsc{HEALPix} maps around the observer at each time step using trilinear interpolation, and then saved to disk.

In Fig.~\ref{fig:HealpixMaps}, we show examples of these \textsc{HEALPix} maps for both the $\delta_{\rm cb}$ and the $\delta_{\nu}$ fields. On the left panel we have these saved for the full-sky light cone at a redshift of $z\approx0.1$, and on the right we show the pencil-beam light cone at a redshift of $z\approx1$.

\paragraph*{Power spectra}---
For each simulation, we also store a set of power spectra for 14 different redshift slices: $z=$ \numlist{50; 30; 10; 4; 3; 2.5; 2; 1.5; 1; 0.75; 0.5; 0.25; 0.1; 0}. Thanks to \textit{gevolution}'s ability to calculate all metric data for the entire simulation, we are able to store power spectra for a multitude of variables. In this case, we extract power spectra for both the scalar and vector gravitational potentials,
as well as for the CDM+baryon density and the total matter density (i.e.\ including the massive neutrinos). This results in $14 \times 4 = 56$ power spectra per simulation, which can be used directly without needing to reanalyse the simulation data.

\begin{figure}
    \centering
    \includegraphics[width=\columnwidth]{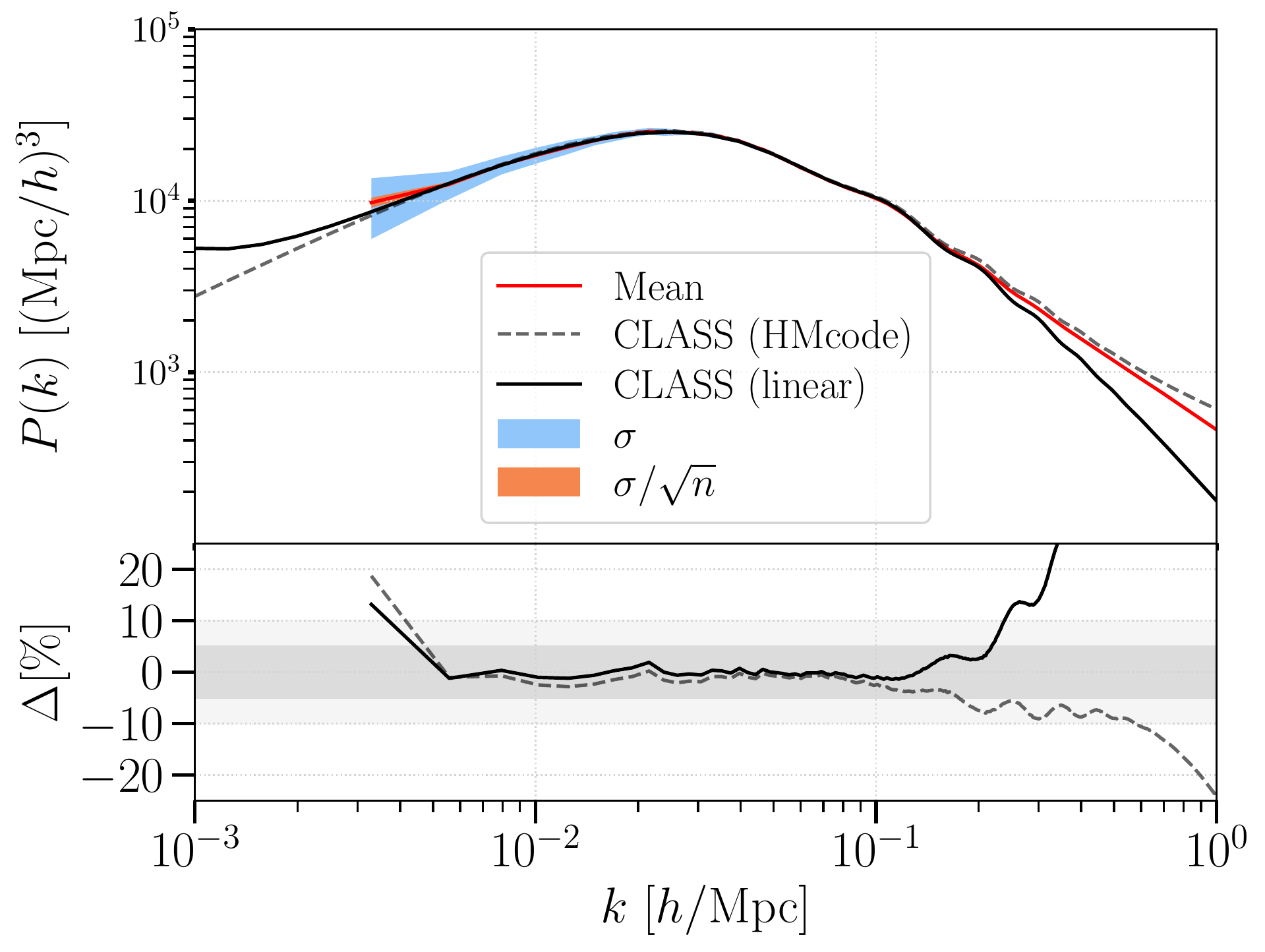}
    \caption{{\it Top panel:} Power spectrum of the CDM+baryon distribution at $z\simeq 0$. The solid red line shows the mean value of the 34 random realisations of the fiducial cosmology, with the standard deviation (and error on the mean) shown as blue and orange 68\% confidence intervals respectively. The black solid and dashed
    lines show theoretical predictions using the \textsc{CLASS} linear power spectrum, and HMcode non-linear model, respectively. HMcode uses the synchronous gauge while the other cases are in Poisson gauge, which explains the different behaviour as one approaches the horizon scale. {\it Bottom panel}: Fractional difference between the theoretical predictions and the mean. The light and dark grey regions show 10\% and 5\% differences between the theory and the mean respectively.}
    \label{fig:PowerSpectra}
\end{figure}

In Fig.~\ref{fig:PowerSpectra} we plot the mean of all of the CDM+baryon power spectra over the 34 random realisations of the fiducial cosmology at $z\simeq 0$, plus confidence intervals showing the expected sample variance (estimated from the simulations) and the error on the mean. We also plot theoretical predictions for comparison, calculated using a linear power spectrum from \textsc{CLASS} \citep*{blas2011} and a non-linear power spectrum model from HMcode \citep{mead2016}. The deviations at large scales are due to the different gauges (Poisson gauge for \textit{gevolution} and the linear prediction of \textsc{CLASS}, while HMcode uses synchronous gauge), while there is good agreement on intermediate and small scales (up to the expected non-linear corrections).

\begin{figure}
    \centering
    \includegraphics[width=\columnwidth]{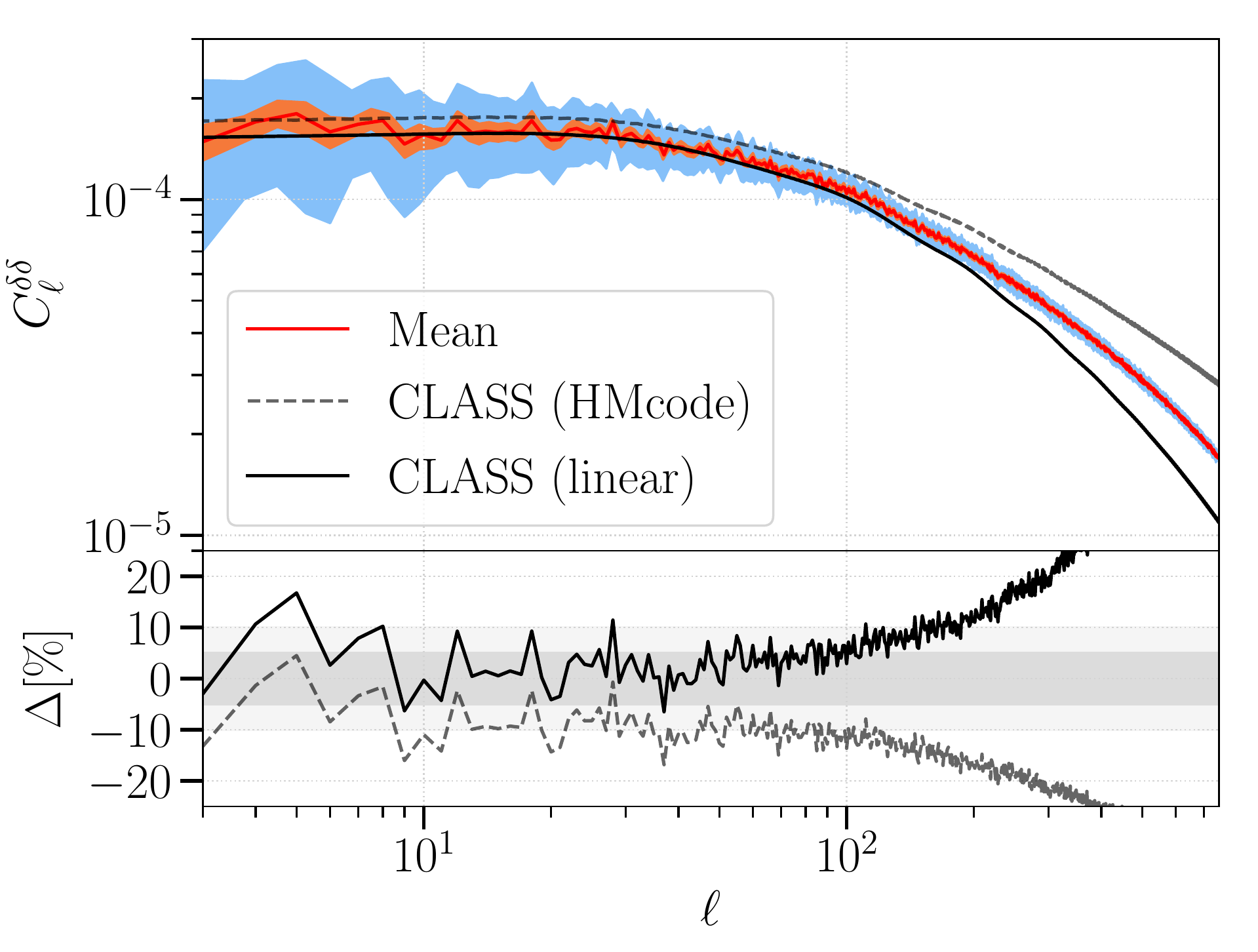}
    \caption{{\it Top panel:} Angular power spectrum of the CDM+baryon field evaluated on the past light cone at $z\approx 0.58$. The mean and standard deviation over the 34 random realisations of the fiducial cosmology are shown as the solid red line and blue shaded region respectively, while the orange region shows the corresponding standard error of the mean (as in Fig.~\ref{fig:PowerSpectra}).  We also show theoretical predictions based on the linear power spectrum for CDM and baryons from \textsc{CLASS}, and the non-linear power spectrum from HMcode, convolved with the CIC kernel (see Sec.~\ref{sec:CrossCorrelation}). {\it Bottom panel:} Fractional difference between the theoretical predictions and the mean. The light and dark grey regions show 10\% and 5\% differences between the theory and the mean respectively.}
    \label{fig:AngPowerSpectra}
\end{figure}

\paragraph*{Angular power spectra}--- 
For completeness, in Fig.~\ref{fig:AngPowerSpectra} we present the angular power spectra computed using full-sky light cones from the 34 random realisations of the fiducial cosmology. We employ the simple quadratic estimator 
\begin{equation}
	\hat{C}_{\ell} = \frac{1}{2\ell + 1}\sum_{m=-\ell}^{\ell} a_{\ell m} a^*_{\ell m},
\end{equation}where $a_{\ell m}$ are the spherical harmonic expansion coefficients of some scalar field $A$ projected onto the unit sphere \citep{Leistedt:2013gfa}. For this example, we took $A = \delta_{\rm cb}$ and performed the analysis using the \textsc{HEALPix} \texttt{anafast} routine, which is suitable for the full-sky case. The shaded region in Fig.~\ref{fig:AngPowerSpectra} corresponds to the standard deviation of $\hat{C}_\ell$ over the 34 samples, with the theoretical angular power spectrum $C_{\ell}$ computed as
\begin{equation}
\label{eq:projection}
	C_{\ell}(z) = \frac{2}{\pi} \int k^2 dk\,\, P_{\rm cb}^s(k,z) \, j^2_{\ell}(k r(z)),
\end{equation}
where $P_{\rm cb}^s(k,z)$ is the power spectrum accounting for baryons and cold dark matter only, smoothed by the cloud-in-cell (CIC) kernel, as discussed in Sec.~\ref{sec:CrossCorrelation}. We projected each map at $z \approx 0.58$, which is equivalent to a comoving distance of $r(z) = 1500$ Mpc/$h$.

In this case, the theoretical calculations are broadly consistent with the measured angular power spectrum, but a slight amplitude offset can be seen, particularly for the non-linear theoretical curve at $\ell \lesssim 100$. From several consistency checks, we have found a contribution coming from the mildly- and fully-nonlinear scales that leaks to large angular scales from the projection integral, Eq.~\eqref{eq:projection}. By comparing the output angular power spectrum from CLASS, for CDM+baryons, the same offset was observed. However, as we can see, both linear and non-linear predictions match the mean value of the simulations at a level of 10\% for $\ell \lesssim 100$. On the other hand, the discrepancies at higher $\ell$ are primarily caused by resolution effects.

\begin{figure*}
    \centering
    \includegraphics[scale=0.48]{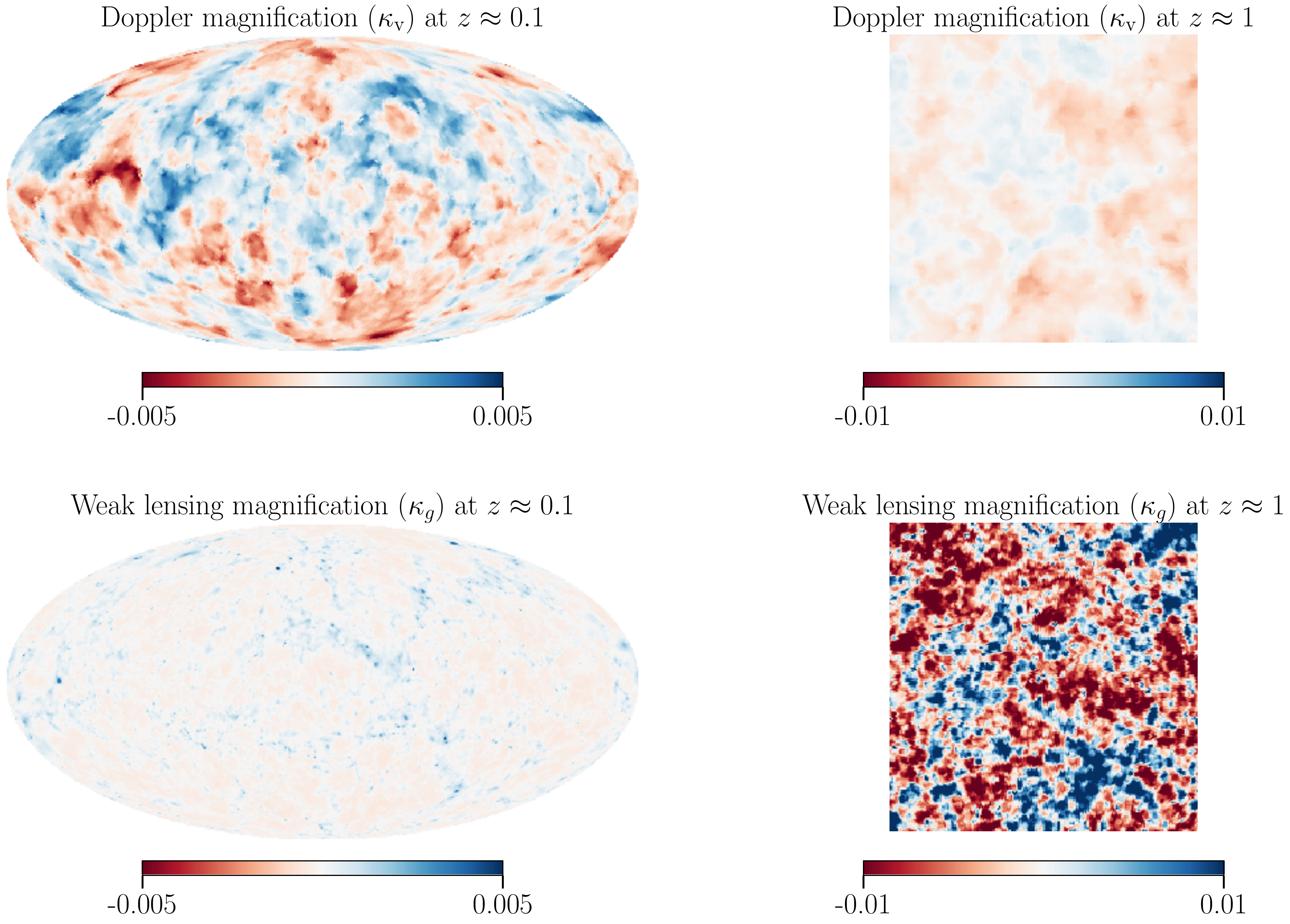}
    \caption{Heat maps for both the Doppler convergence and the weak-lensing convergence. We show the full-sky signal at  redshift $z\approx0.1$ and the pencil beam at $z\approx 1$, plotting a $10^\circ \times 10^\circ$ section. For each redshift we keep the scale of the colour bars the same to show the difference between the two signals.}
    \label{fig:HealpixMapsLensing}
\end{figure*}

\paragraph*{Ray-traced quantities}--- 
Given a set of \textsc{HEALPix} maps of the gravitational potential $\phi$ on the light cone, it is straightforward to construct other interesting quantities that are linearly related to $\phi$ or its time derivative. Examples include the integrated Sachs-Wolfe effect and the weak-lensing potential. Using the Born approximation, \textsc{HEALPix} maps of these quantities can be computed directly in pixel space by adding together maps of $\phi$ with appropriate weights. For instance, the weak-lensing potential, which is defined in \citet{Lewis:2006} as
\begin{equation}
    \Psi(\boldsymbol{\theta}, z) \equiv-\int_{0}^{r(z)} d r^{\prime} \, \frac{r(z)-r^{\prime}}{r(z) r^{\prime}}(\phi+\psi)\,,
    \label{eq:lensing_potential}
\end{equation}
can be constructed by such a procedure, explained in more detail in \citet{Lepori:2020ifz}. Here we make the assumption that $\psi \approx \phi$, which is an excellent approximation here, and neglect the effect of frame dragging.
The HEALPix maps can then be easily converted into linear weak-lensing convergence and shear maps by using
\begin{align}
\kappa_g &=-\frac{1}{2} \Delta \Psi\,, \\
\gamma_{1}+i \gamma_{2} &=-\frac{1}{2}\left(\nabla_{1} \nabla_{1}-\nabla_{2} \nabla_{2}\right) \Psi-\mathrm{i} \nabla_{1} \nabla_{2} \Psi\,,
\end{align}
where $\Delta$ is the Laplacian, and the derivatives are taken on the map. The convergence $\kappa_g$ (we use the subscript ``$g$'' to distinguish this term from Doppler magnification, as discussed below) and the shear $\gamma$ parametrise the amplification matrix,
\begin{equation}
\label{eq:amplificationmatrix}
    A=\left(\begin{array}{cc}
    1-\kappa_g-\gamma_{1} & -\gamma_{2} \\
    -\gamma_{2} & 1-\kappa_g+\gamma_{1}
    \end{array}\right),
\end{equation}
that relates lensed images to unlensed ones if lensing is treated linearly \citep{Bartelmann:1999yn}. It is worth pointing out that non-linear ray tracing is also possible with the data, e.g.\ using the full machinery developed in \citet{Lepori:2020ifz}, although we do not pursue this here.


\section{Results}\label{sec:Results}

In this section, we show one of the many ways these simulations can be used to construct relativistic observables by using the {\it Doppler magnification} effect as an example. This was first highlighted and investigated as an observable in its own right by \citet{Bonvin:2005ps, Bonvin:2008ni}. Later works have shown that this signal should be detectable with modern day optical and radio surveys, \cite[e.g.][]{Bonvin:2016dze,Andrianomena:2018aad}.

A perturbative expression was derived for the cross-correlation between the Doppler magnification signal and the matter density in \citet{Bacon:2014uja}. Doppler magnification is a relativistic effect that is caused by the relative motion of the source and the observer, which correlates with matter density as sources will tend to fall towards areas of high density.
It is an inherently (special) relativistic effect, and while it is also possible to derive it from Newtonian simulations, this requires a more careful handling of gauge issues etc. than is needed here. To illustrate their utility, we will go through the steps in calculating this signal within our suite of simulations and then compare with the perturbative results from \citet{Bacon:2014uja}.

\subsection{Doppler magnification}
\label{sec:DopplerLensing}

In the linear weak-lensing regime, the true shape of the source is related to the observed image through the Jacobi map,
\begin{equation}
 \mathcal{J} = \bar{D}_A(r) A\,,
\end{equation}
where $\bar{D}_A$ is the angular diameter distance to the source in the background metric, and $A$ is the amplification matrix given in Eq.~\eqref{eq:amplificationmatrix}. Doppler magnification appears if one uses the observed redshift as a distance indicator, i.e.\ the Jacobi map is written at fixed observed redshift as
\begin{align}
    \mathcal{J} &= \bar{D}_A(z_s) \left(1 - \frac{\partial \ln\bar{D}_A}{\partial z} \delta z\right) A \nonumber \\ &\simeq \bar{D}_A(z_s) \left(\begin{array}{cc}
    1-\kappa_g - \kappa_v-\gamma_{1} & -\gamma_{2} \\
    -\gamma_{2} & 1-\kappa_g - \kappa_v +\gamma_{1}
    \end{array}\right),
\end{align}
where one defines

\begin{equation}
    \kappa_{v}=\frac{\partial \ln\bar{D}_A}{\partial z} \delta z =\left(\frac{1+z_{s}}{H \, r_{s}} - 1\right) \boldsymbol{v}_{s} \cdot \boldsymbol{n}\,.
    \label{eq:doppler_lensing}
\end{equation}
Here $\boldsymbol{v}_{s}$ is the peculiar velocity of the source, and $z_{s}$ and $r_{s}$ are the redshift and the comoving radial distance to the source, respectively. We define the direction of $\boldsymbol{n}$ to be pointed from the observer to the source.

It can be seen that galaxies that have a velocity vector directed towards the observer will create a negative $\kappa_{v}$ at low redshift, which means that they will appear smaller in angular size and dimmer when compared to a typical source at the same observed redshift. In contrast, if the source is moving away from the observer, the $\kappa_{v}$ will be positive, which means it will appear brighter with a larger angular size. This effect comes about because a surface of fixed observed redshift does not coincide with a surface of fixed comoving distance. Here and in the following, we assume that the Doppler contribution from the source peculiar motion is the only relevant redshift perturbation, i.e.\ we neglect gravitational redshift and other subdominant corrections. Note that all such corrections {\it can} be included by combining the appropriate fields when ray-tracing, however.

\begin{figure*}
    \centering
    \includegraphics[scale=0.63]{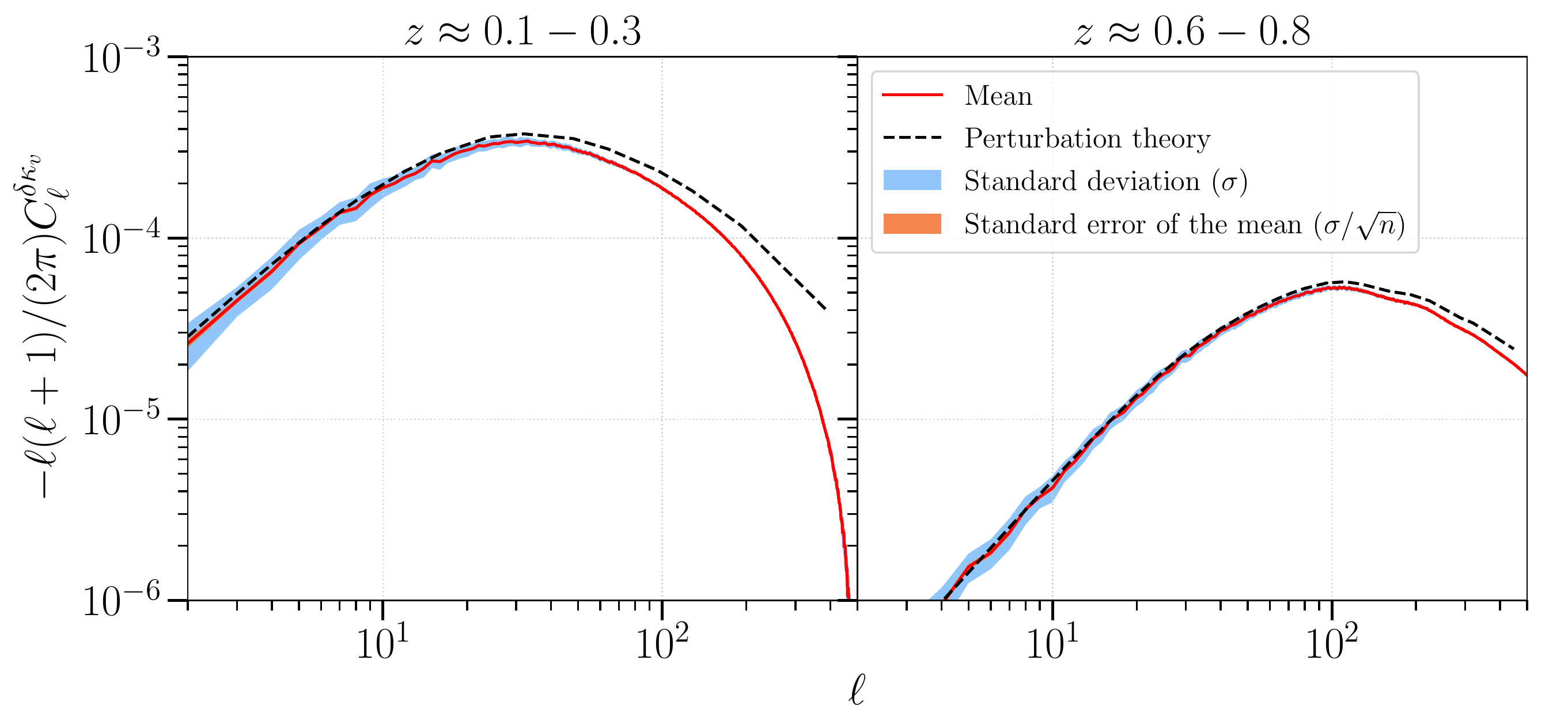}
    \caption{Plot of the mean $C_{\ell}^{\delta \kappa_{v}}$ across all of the different random realisations of the fiducial cosmology for two different redshift bins. We also plot the standard deviation and the standard error of the mean which are represented by the shaded areas. The dashed line on the plot shows the theoretical prediction of the signal from Eq.~\eqref{eq:cltheory}.}
    \label{fig:mean}
\end{figure*}

Another important point to note from Eq.~\eqref{eq:doppler_lensing} is that at high redshifts the term in brackets will decrease, causing a lower amplitude of Doppler magnification. Using this fact and the estimate $|\boldsymbol{v}| \sim H_{0} \delta / k$, one expects that the Doppler magnification is only important on large scales and at low redshift \citep{Bolejko:2012uj, Bacon:2014uja}. Fig.~2 of \citet{Bacon:2014uja} does indeed show that the Doppler magnification term is dominant at medium-to-low redshifts and wavenumbers ($\ell \lesssim 1000$ at $z = 0.2$, and $l \lesssim 100$ at $z = 0.4$).

The relative importance of the different contributions to the observed convergence can also be judged from Fig.~\ref{fig:HealpixMapsLensing}, where we show the Doppler magnification signal and the weak lensing signal within our simulation. To calculate the Doppler magnification signal we use Eq.~\eqref{eq:doppler_lensing} together with maps of the redshift space distortion field that are included in our data products.
For the full-sky maps, we show the signal at $z\approx0.1$, where it can be seen that the Doppler magnification is much stronger than the weak lensing signal. When we look at the pencil beam at $z\approx2$, we see that the weak-lensing effect dominates instead, as this integrated effect becomes stronger with increasing distance.

\subsection{Density-convergence cross-correlation}\label{sec:CrossCorrelation}

\begin{figure*}
    \centering
    \includegraphics[scale=0.5]{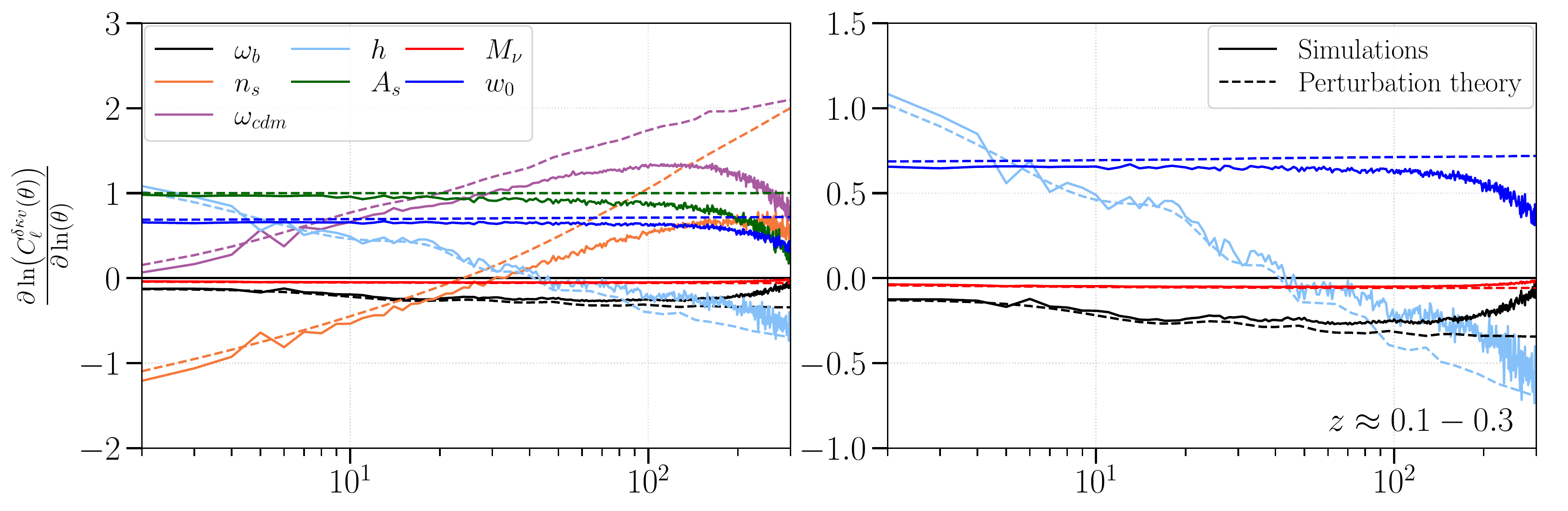}
    \includegraphics[scale=0.5]{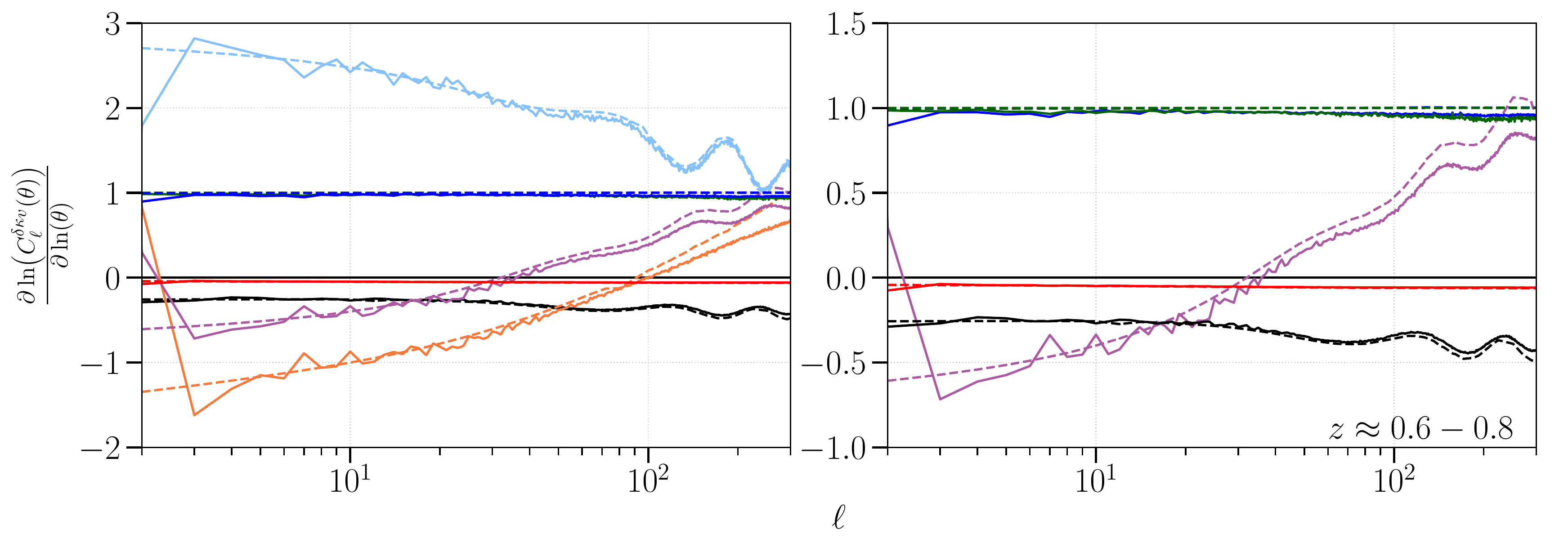}
    \caption{Plots of the numerical derivatives of $C_\ell^{\delta \kappa_v}$ using finite differences of the density Doppler magnification cross correlation signal in the redshift bin $z \approx 0.1-0.3$ ({\it upper panels}) and $z \approx 0.6-0.8$ ({\it lower panels}). This is done for 7 pairs of simulations where $\theta = A_s$, $n_s$, $h$, $\omega_{cdm}$, $\omega_b$, $M_\nu$ and $w_0$ are varied. The solid lines show the results measured from the simulations whereas the dashed lines show the perturbative predictions. The right panels are zoomed-in sections of the left panels so the difference in the selected quantities can be seen more clearly.}
    \label{fig:finite_diff} \label{fig:finite_diff_high}
\end{figure*}

Since matter tends to collapse onto massive structures, there is an obvious correlation between the Doppler magnification and the density field. At low redshift, the Doppler magnification of sources on the far side of a large concentration of matter tends to be negative, while the opposite is true for sources on the near side. In order to measure this correlation, we take the average density over a small interval in distance from the observer, $[r, r + \Delta r]$, and compute the angular cross-power with the Doppler convergence, $\kappa_v$, evaluated at the far end of the interval. We then compute the average of the resulting cross-power over a larger distance range in order to accumulate a larger total signal.

A full derivation of the angular cross-correlation from perturbation theory is presented in \citet{Bacon:2014uja}. We present only the final result in Poisson gauge here. We obtain
\begin{multline}
    C_{\ell}^{\delta \kappa_{v}}\left( r^\prime \right) = \frac{16 \pi^2}{\Delta r}\left(H\left( r^\prime\right)-\frac{1+z\left( r^\prime\right)}{ r^\prime}\right)\frac{\partial D\left( r^\prime\right)}{\partial z}\\
    \times \int_{0}^{\infty} \mathrm{d} k \,\, P_{\rm cb}(k) \, k \, j^{\prime}_{\ell}\left(k  r^\prime\right) \\
    \times \int_{ r^{\prime}-\Delta r}^{ r^{\prime }} \mkern-18mu \mathrm{d} r \left(D\left( r \right)- \left(\frac{3H( r)^2}{\left(1+z\left( r\right)\right)k^2}\right)\frac{\partial D\left( r\right)}{\partial z} \right) j_{\ell}(k  r),
\label{eq:cltheory}
\end{multline}where $P_{\rm cb}(k)$ is the CDM + baryon power spectrum at redshift zero and $D(r)$ is the linear growth factor,\footnote{In Eq.~(\ref{eq:cltheory}), scale-independent growth has been assumed for simplicity. This is a good approximation on scales much smaller than the neutrino free-streaming scale (and in the case of $w_0 < -1$, the sound horizon scale of dark energy perturbations), where $D$ is computed neglecting any perturbations in components other than CDM and baryons. Scale-dependent growth is fully taken into account in the simulations themselves.} defined as $\delta(k, r) = D(r)\delta(k, 0)$. Note that, in a slight variation to \citet{Bacon:2014uja}, we employ weights that are uniform in comoving distance $r$ instead of weights that are uniform in redshift.

Matter overdensities only have an appreciable gravitational influence over short distances, expected to be somewhere in the region of tens of megaparsecs. Without trying to make an optimal choice, we set $\Delta  r = 52.5$ Mpc/$h$ for the distance window in which the density is computed, ignoring longer-range correlations. 
The cross-correlation signal $C_{\ell}^{\delta \kappa_{v}}(r^\prime)$ can then be averaged over a broader bin to get the average angular cross-correlation within the bin, which we will denote as $C_{\ell}^{\delta \kappa_{v}}$.

To compute this value in our simulations we use the \textsc{HEALPix} maps of the line-of-sight peculiar velocity field $\boldsymbol{v}_{s} \cdot \boldsymbol{n}$ and the CDM+baryon density as described in Sec.~\ref{sec:DataProducts}.
Specifically, for each radial shell we use maps from the two consecutive simulation time steps that together enclose the light cone at the given distance and
\%For each of these radial shells, we save two additional shells at time steps on either side. This allows us to 
calculate the data on the null hypersurface by linear interpolation in conformal time. From these data, we create a thin bin of $\Delta  r = 52.5$ Mpc/$h$ where we sum up the density maps, and then cross-correlate with the Doppler magnification map directly on the far side of the bin, which we compute from Eq.~\eqref{eq:doppler_lensing}. The cross-correlation is calculated using the \textsc{HEALPix} \texttt{anafast} function. We then repeat for all thin bins within the thick bin and take the average to obtain $C_{\ell}^{\delta \kappa_{v}}$.

To accurately compare the simulations to the perturbative prediction, it is necessary to convolve the power spectrum in the perturbative calculation with the cloud-in-cell (CIC) kernel \citep{1981csup.book.....H}. This effectively gives a smoothed power spectrum which accounts for the fact that the density field in the simulations is coarse-grained at a finite resolution.
The expression for the monopole of this smoothed power spectrum is
\begin{equation}
    P^s_0\left(k\right) = P(k)\, \frac{1}{4\pi} \int \mathrm{d}^2 n \,\, W^2_{\rm CIC}\left(k,\boldsymbol{n}\right),
    \label{eq:smoothed_power}
\end{equation}
where the CIC kernel $W_{\rm CIC}(\boldsymbol{k})$ is defined as
\begin{equation}
    W_{\rm CIC}(\boldsymbol{k})=
    \mathrm{sinc}^2 \left(\frac{\pi k_{1}}{2 k_{N}}\right) \mathrm{sinc}^2 \left(\frac{\pi k_{2}}{2 k_{N}}\right)  \mathrm{sinc}^2 \left(\frac{\pi k_{3}}{2 k_{N}}\right)\,,
    \label{eq:CICKernal}
\end{equation}
where $k_i$ is the $i$-th component of $\boldsymbol{k}$ and $k_N$ is the Nyquist wavenumber. The smoothed monopole power spectrum of Eq.~\eqref{eq:smoothed_power} is then substituted for $P_{\rm cb}$ in Eq.~\eqref{eq:cltheory}.

In Fig.~\ref{fig:mean} we show the distribution of $C_{\ell}^{\delta \kappa_{v}}$ measured from all realisations of the fiducial cosmology, with the mean shown as a solid red line and the standard deviation shown as a blue shaded region. We also plot the perturbative prediction calculated from Eq.~\eqref{eq:cltheory} as a black dashed line. The results are shown for two redshift bins, the first at $z \approx 0.1-0.3$, and the second at $z \approx 0.6-0.8$.

In the lower redshift bin, perturbation theory overestimates the signal beyond $\ell \gtrsim 30$ or so. This difference is mostly due to non-linear effects on small and intermediate scales that are not included in the linear power spectrum model, caused by orbit crossings that generate both vorticity and velocity dispersion and at the same time reduce the power in the velocity divergence 
\citep{pueblas2009, Hahn_2015, Jelic_Cizmek_2018}. As expected, this effect is also present in the density-velocity cross-correlation. Since this bin is at relatively low redshift, non-linear effects are important even at quite low values of $\ell$. In the higher redshift bin, on the other hand, the value measured from our simulations fits more closely to the perturbative prediction, although the strength of the signal has decreased by around an order of magnitude by this point. Also note the shift of the peak in the cross-correlation to correspondingly smaller angular scales.

In Fig.~\ref{fig:finite_diff} we show the numerical derivatives of $C_\ell^{\delta \kappa_v}$ by using finite differences (Eq.~\eqref{eq:finite_diff}) of this quantity in the two redshift bins, $z \approx 0.1-0.3$ (upper panels) and $z \approx 0.6-0.8$ (lower panels), for seven pairs of simulations with the same initial conditions but different values of the cosmological parameters $A_s$, $n_s$, $h$, $\omega_{cdm}$, $\omega_b$, $M_\nu$ and $w_0$ as described in Table~\ref{tab:simulations}. In the case of varying $w_0$, the pair of simulations used are the baseline cosmology and the one where we include perturbations in the dark energy field (see Sec.~\ref{sec:UnitySuite}). 
 Derivatives of this kind are useful for Fisher forecasting studies, and give a direct measure of the sensitivity of an observable to a given parameter.

Theoretical predictions of the derivatives from perturbation theory are also shown as dashed lines in Fig.~\ref{fig:finite_diff}. As in previous figures, we see differences of a few percent between the perturbation theory and simulated quantities in the lowest redshift bin (upper panels), which is consistent with the growing importance of non-linear effects at these redshifts, as discussed above. At higher redshift (lower panels), these effects are subdominant however, and the agreement between perturbation theory and the simulated quantities is good to within a couple of percent even on smaller angular scales.

A notable feature of Fig.~\ref{fig:finite_diff} is the wiggle-like features in some of the curves for the higher redshift bin. These are due to shifts in the location of the baryon acoustic oscillations, which could also be seen (albeit at quite a low level) in Fig.~\ref{fig:mean}. While the signal-to-noise ratio of any practical measurement of the Doppler magnification signal is unlikely to be sufficient to detect the BAO feature for the foreseeable future, it is interesting that they can in principle be picked up by this observable.

\section{Conclusions}\label{sec:Conclusions}

In this paper, we have presented a suite of novel general relativistic cosmological simulations. These were run using the \textit{gevolution} $N$-body code \citep{Adamek:2016zes}, which takes into account all relevant relativistic effects, including frame dragging and relativistic neutrino effects. While most of those effects are small in comparison with the conventional `Newtonian' terms in most large-scale observables, the rapidly increasing precision of upcoming surveys will soon make it impractical to ignore them without risking biases in cosmological parameter estimates.

The suite of simulations contains a total of 53 runs, divided into subsets that are envisioned to have two main uses. The first subset contains 39 simulations that all use the same fiducial cosmology, but vary the random seed used to generate the initial conditions, which essentially gives us different realisations of the same underlying cosmology. Possible applications of this subset include statistical studies of observables, estimators, and data extraction methods, plus some rudimentary kinds of simulation-based covariance estimation.

The second part of this suite consists of 7 pairs of simulations with cosmological parameters that are systematically varied around the fiducial cosmology (which matches the Euclid Flagship 2 cosmology), while maintaining the same (random) initial conditions. These allow us to study the derivatives of any observable that we can calculate with respect to a set of cosmological parameters. Possible applications of this suite include Fisher forecasting, where derivatives of observables are used to estimate the uncertainties on measurements that can be achieved by future experiments.

We have stored a range of data products for each simulation, including all of the metric degrees of freedom and other fields needed to reconstruct any cosmological observable on large scales. These fields have been determined in a spherical coordinate system about a fiducial observer, and can be fed into a ray-tracing algorithm to produce precise predictions of observables on the past light cone. The geometry of the simulations has been chosen to maximise the sky area and depth of the light cones that can be simulated, with the full sky accessible out to $z=0.85$ and a large area (1930 sq. deg.) available out to $z=3.55$. These specifications are well-matched to a variety of current and near-future large-scale structure surveys, including the ESA Euclid mission, the Roman Space Telescope, the VRO Legacy Survey of Space and Time, and the Square Kilometre Array. While the simulations do not have sufficient resolution to produce suitable dark matter halo catalogues for these surveys, biased tracers can be painted onto the simulations using other means \citep[e.g.][]{Borzyszkowski:2017ayl, Bull:2016ziy, Witzemann:2018cdx, Yip:2019rfl, Ramanah:2020vyl, Farr:2019xij}

To showcase the potential use-cases of our suite of simulations, we calculate the cross-correlation of the inherently special relativistic Doppler magnification signal, $\kappa_{v}$, and the matter density contrast, $\delta_{\rm cb}$, from our 34 random realisation simulations and compare it to the perturbation theory result from \citet{Bacon:2014uja}. We find good agreement with the perturbation theory calculation in a relatively high redshift bin of $z \approx 0.6-0.8$, but find non-negligible corrections in a lower redshift bin of $z \approx 0.1-0.3$, where linear theory overestimates the signal. This is due to non-linear effects that appear on small scales which are not included in the linear prediction, and which (due to projection effects) affect most of the relevant angular scales at the low redshifts where the Doppler magnification signal is largest. While an improvement, replacing the linear matter power spectrum in the perturbation theory calculation with a non-linear power spectrum model does not fully capture these effects (c.f. Figs.~\ref{fig:PowerSpectra} and \ref{fig:AngPowerSpectra}). This highlights the value of having fully relativistic cosmological simulations on hand to make predictions of such observables.

We also calculated the Doppler magnification cross-correlation signal for matched pairs of simulations with the same initial conditions but different values for the cosmological parameters. This allowed us to approximate the derivatives of the observable $C_{\ell}^{\delta \kappa_{v}}$ and therefore see which parameters it is most sensitive to. While we again saw generally good agreement with predictions from perturbation theory, especially in the higher redshift bin, non-negligible corrections remained. Since reasonably any large-scale structure observable can be constructed from our suite of simulations, including many different combinations of cross-correlations and even high-order statistics, it should be possible to make Fisher matrix-type forecasts for a very wide range of surveys and observables using these data.

In conclusion, in this paper we have described a suite of fully relativistic $N$-body simulations, and shown a particular example (the Doppler magnification term in the density-convergence cross-correlation) in which such simulations are needed in order to make accurate predictions for next-generation surveys. While perturbation theory calculations were able to capture most features of the target signal, non-linear effects made few-percent differences at low redshifts. To accurately model these in perturbation theory, one would likely have to include higher-order corrections in redshift-space which are difficult to compute.

\vspace{-2em}
\section*{Acknowledgements}

This work (project {\tt dp103}) used the DiRAC DIaL system, operated by the University of Leicester IT Services, which forms part of the STFC DiRAC HPC Facility (\url{www.dirac.ac.uk}). This equipment is funded by BIS National E-Infrastructure capital grant ST/K000373/1 and STFC DiRAC Operations grant ST/K0003259/1. DiRAC is part of the National e-Infrastructure. In addition, this work was supported by a grant from the Swiss National Supercomputing Centre (CSCS) under project ID s710. PB acknowledges support from STFC grant ST/T000341/1. JA acknowledges financial support by STFC Consolidated Grant ST/P000592/1 and by the Swiss National Science Foundation. CG is supported by FAPESP (\href{https://bv.fapesp.br/pt/bolsas/180562/tecnicas-estatisticas-para-levantamentos-futuros-extraindo-fisica-primordial-da-estrutura-em-larga/?q=2018/10396-2}{2018/10396-2}) grant. Some of the results in this paper have been derived using the healpy and HEALPix packages \citep{2005ApJ...622..759G, Zonca2019}. 

\vspace{-2em}
\section*{Data availability}
The data that supports the findings of this research is available on request from \href{http://philbull.com/unity}{http://philbull.com/unity}.

\vspace{-2em}
\section*{Carbon footprint}
The numerical simulations presented in this article used about $3900~\mathrm{kWh}$ of electrical energy. Using a conversion factor of $0.681~\mathrm{kg\,CO_2\,kWh^{-1}}$ (typical for the UK grid according to \href{https://co2.myclimate.org/en}{myclimate.org}, c. 12\ Feb.~2021) this gives a carbon footprint of approximately $2.7~\mathrm{t\,CO_2}$.


\bibliographystyle{mnras}
\bibliography{refs.bib}

\begin{thebibliography}{}
\makeatletter
\relax
\def\mn@urlcharsother{\let\do\@makeother \do\$\do\&\do\#\do\^\do\_\do\%\do\~}
\def\mn@doi{\begingroup\mn@urlcharsother \@ifnextchar [ {\mn@doi@}
  {\mn@doi@[]}}
\def\mn@doi@[#1]#2{\def\@tempa{#1}\ifx\@tempa\@empty \href
  {http://dx.doi.org/#2} {doi:#2}\else \href {http://dx.doi.org/#2} {#1}\fi
  \endgroup}
\def\mn@eprint#1#2{\mn@eprint@#1:#2::\@nil}
\def\mn@eprint@arXiv#1{\href {http://arxiv.org/abs/#1} {{\tt arXiv:#1}}}
\def\mn@eprint@dblp#1{\href {http://dblp.uni-trier.de/rec/bibtex/#1.xml}
  {dblp:#1}}
\def\mn@eprint@#1:#2:#3:#4\@nil{\def\@tempa {#1}\def\@tempb {#2}\def\@tempc
  {#3}\ifx \@tempc \@empty \let \@tempc \@tempb \let \@tempb \@tempa \fi \ifx
  \@tempb \@empty \def\@tempb {arXiv}\fi \@ifundefined
  {mn@eprint@\@tempb}{\@tempb:\@tempc}{\expandafter \expandafter \csname
  mn@eprint@\@tempb\endcsname \expandafter{\@tempc}}}

\bibitem[\protect\citeauthoryear{Abell et~al.}{Abell
  et~al.}{2009}]{Abell:2009aa}
Abell P.~A.,  et~al., 2009, arXiv:0912.0201

\bibitem[\protect\citeauthoryear{Adamek, Daverio, Durrer  \& Kunz}{Adamek
  et~al.}{2016a}]{Adamek:2016zes}
Adamek J.,  Daverio D.,  Durrer R.,   Kunz M.,  2016a, \mn@doi [JCAP]
  {10.1088/1475-7516/2016/07/053}, 07, 053

\bibitem[\protect\citeauthoryear{Adamek, Daverio, Durrer  \& Kunz}{Adamek
  et~al.}{2016b}]{Adamek:2015eda}
Adamek J.,  Daverio D.,  Durrer R.,   Kunz M.,  2016b, \mn@doi [Nature Phys.]
  {10.1038/nphys3673}, 12, 346

\bibitem[\protect\citeauthoryear{Alonso, Bull, Ferreira, Maartens  \&
  Santos}{Alonso et~al.}{2015}]{Alonso:2015uua}
Alonso D.,  Bull P.,  Ferreira P.~G.,  Maartens R.,   Santos M.,  2015, \mn@doi
  [Astrophys. J.] {10.1088/0004-637X/814/2/145}, 814, 145

\bibitem[\protect\citeauthoryear{Andrianomena, Bonvin, Bacon, Bull, Clarkson,
  Maartens  \& Moloi}{Andrianomena et~al.}{2019}]{Andrianomena:2018aad}
Andrianomena S.,  Bonvin C.,  Bacon D.,  Bull P.,  Clarkson C.,  Maartens R.,
  Moloi T.,  2019, \mn@doi [Mon. Not. Roy. Astron. Soc.]
  {10.1093/mnras/stz1905}, 488, 3759

\bibitem[\protect\citeauthoryear{Bacon, Andrianomena, Clarkson, Bolejko  \&
  Maartens}{Bacon et~al.}{2014}]{Bacon:2014uja}
Bacon D.~J.,  Andrianomena S.,  Clarkson C.,  Bolejko K.,   Maartens R.,  2014,
  \mn@doi [Mon. Not. Roy. Astron. Soc.] {10.1093/mnras/stu1270}, 443, 1900

\bibitem[\protect\citeauthoryear{Bartelmann \& Schneider}{Bartelmann \&
  Schneider}{2001}]{Bartelmann:1999yn}
Bartelmann M.,  Schneider P.,  2001, \mn@doi [Phys. Rept.]
  {10.1016/S0370-1573(00)00082-X}, 340, 291

\bibitem[\protect\citeauthoryear{Bertacca, Maartens, Raccanelli  \&
  Clarkson}{Bertacca et~al.}{2012}]{Bertacca:2012tp}
Bertacca D.,  Maartens R.,  Raccanelli A.,   Clarkson C.,  2012, \mn@doi [JCAP]
  {10.1088/1475-7516/2012/10/025}, 10, 025

\bibitem[\protect\citeauthoryear{Bertacca, Raccanelli, Bartolo, Liguori,
  Matarrese  \& Verde}{Bertacca et~al.}{2018}]{Bertacca:2017dzm}
Bertacca D.,  Raccanelli A.,  Bartolo N.,  Liguori M.,  Matarrese S.,   Verde
  L.,  2018, \mn@doi [Phys. Rev. D] {10.1103/PhysRevD.97.023531}, 97, 023531

\bibitem[\protect\citeauthoryear{Beutler \& Di~Dio}{Beutler \&
  Di~Dio}{2020}]{Beutler:2020evf}
Beutler F.,  Di~Dio E.,  2020, \mn@doi [JCAP] {10.1088/1475-7516/2020/07/048},
  07, 048

\bibitem[\protect\citeauthoryear{Blas, Lesgourgues  \& Tram}{Blas
  et~al.}{2011}]{blas2011}
Blas D.,  Lesgourgues J.,   Tram T.,  2011, \mn@doi [JCAP]
  {10.1088/1475-7516/2011/07/034}, 2011, 034

\bibitem[\protect\citeauthoryear{Bolejko, Clarkson, Maartens, Bacon, Meures  \&
  Beynon}{Bolejko et~al.}{2013}]{Bolejko:2012uj}
Bolejko K.,  Clarkson C.,  Maartens R.,  Bacon D.,  Meures N.,   Beynon E.,
  2013, \mn@doi [Phys. Rev. Lett.] {10.1103/PhysRevLett.110.021302}, 110,
  021302

\bibitem[\protect\citeauthoryear{Bonvin}{Bonvin}{2008}]{Bonvin:2008ni}
Bonvin C.,  2008, \mn@doi [Phys. Rev. D] {10.1103/PhysRevD.78.123530}, 78,
  123530

\bibitem[\protect\citeauthoryear{Bonvin}{Bonvin}{2014}]{Bonvin:2014owa}
Bonvin C.,  2014, \mn@doi [Class. Quant. Grav.]
  {10.1088/0264-9381/31/23/234002}, 31, 234002

\bibitem[\protect\citeauthoryear{Bonvin \& Durrer}{Bonvin \&
  Durrer}{2011}]{Bonvin:2011bg}
Bonvin C.,  Durrer R.,  2011, \mn@doi [Phys. Rev. D]
  {10.1103/PhysRevD.84.063505}, 84, 063505

\bibitem[\protect\citeauthoryear{Bonvin, Durrer  \& Gasparini}{Bonvin
  et~al.}{2006}]{Bonvin:2005ps}
Bonvin C.,  Durrer R.,   Gasparini M.,  2006, \mn@doi [Phys. Rev. D]
  {10.1103/PhysRevD.85.029901}, 73, 023523

\bibitem[\protect\citeauthoryear{Bonvin, Andrianomena, Bacon, Clarkson,
  Maartens, Moloi  \& Bull}{Bonvin et~al.}{2017}]{Bonvin:2016dze}
Bonvin C.,  Andrianomena S.,  Bacon D.,  Clarkson C.,  Maartens R.,  Moloi T.,
   Bull P.,  2017, \mn@doi [Mon. Not. Roy. Astron. Soc.]
  {10.1093/mnras/stx2049}, 472, 3936

\bibitem[\protect\citeauthoryear{Borzyszkowski, Bertacca  \&
  Porciani}{Borzyszkowski et~al.}{2017}]{Borzyszkowski:2017ayl}
Borzyszkowski M.,  Bertacca D.,   Porciani C.,  2017, \mn@doi [Mon. Not. Roy.
  Astron. Soc.] {10.1093/mnras/stx1423}, 471, 3899

\bibitem[\protect\citeauthoryear{Brandbyge \& Hannestad}{Brandbyge \&
  Hannestad}{2009}]{Brandbyge:2008js}
Brandbyge J.,  Hannestad S.,  2009, \mn@doi [JCAP]
  {10.1088/1475-7516/2009/05/002}, 05, 002

\bibitem[\protect\citeauthoryear{Bruni, Crittenden, Koyama, Maartens, Pitrou
  \& Wands}{Bruni et~al.}{2012}]{Bruni:2011ta}
Bruni M.,  Crittenden R.,  Koyama K.,  Maartens R.,  Pitrou C.,   Wands D.,
  2012, \mn@doi [Phys. Rev. D] {10.1103/PhysRevD.85.041301}, 85, 041301

\bibitem[\protect\citeauthoryear{Bull}{Bull}{2017}]{Bull:2016ziy}
Bull P.,  2017, \mn@doi [Mon. Not. Roy. Astron. Soc.] {10.1093/mnras/stx1052},
  471, 12

\bibitem[\protect\citeauthoryear{Camera, Maartens  \& Santos}{Camera
  et~al.}{2015}]{Camera:2014sba}
Camera S.,  Maartens R.,   Santos M.~G.,  2015, \mn@doi [Mon. Not. Roy. Astron.
  Soc.] {10.1093/mnrasl/slv069}, 451, L80

\bibitem[\protect\citeauthoryear{Chisari \& Zaldarriaga}{Chisari \&
  Zaldarriaga}{2011}]{Chisari:2011iq}
Chisari N.~E.,  Zaldarriaga M.,  2011, \mn@doi [Phys. Rev. D]
  {10.1103/PhysRevD.84.089901}, 83, 123505

\bibitem[\protect\citeauthoryear{Clarkson, de Weerd, Jolicoeur, Maartens  \&
  Umeh}{Clarkson et~al.}{2019}]{Clarkson:2018dwn}
Clarkson C.,  de Weerd E.~M.,  Jolicoeur S.,  Maartens R.,   Umeh O.,  2019,
  \mn@doi [Mon. Not. Roy. Astron. Soc.] {10.1093/mnrasl/slz066}, 486, L101

\bibitem[\protect\citeauthoryear{Dakin, Hannestad, Tram, Knabenhans  \&
  Stadel}{Dakin et~al.}{2019}]{Dakin:2019vnj}
Dakin J.,  Hannestad S.,  Tram T.,  Knabenhans M.,   Stadel J.,  2019, \mn@doi
  [JCAP] {10.1088/1475-7516/2019/08/013}, 08, 013

\bibitem[\protect\citeauthoryear{De~Weerd, Clarkson, Jolicoeur, Maartens  \&
  Umeh}{De~Weerd et~al.}{2020}]{deWeerd:2019cae}
De~Weerd E.~M.,  Clarkson C.,  Jolicoeur S.,  Maartens R.,   Umeh O.,  2020,
  \mn@doi [JCAP] {10.1088/1475-7516/2020/05/018}, 05, 018

\bibitem[\protect\citeauthoryear{Farr et~al.}{Farr et~al.}{2020}]{Farr:2019xij}
Farr J.,  et~al., 2020, \mn@doi [JCAP] {10.1088/1475-7516/2020/03/068}, 03, 068

\bibitem[\protect\citeauthoryear{Fidler, Tram, Rampf, Crittenden, Koyama  \&
  Wands}{Fidler et~al.}{2017}]{Fidler:2017pnb}
Fidler C.,  Tram T.,  Rampf C.,  Crittenden R.,  Koyama K.,   Wands D.,  2017,
  \mn@doi [JCAP] {10.1088/1475-7516/2017/12/022}, 12, 022

\bibitem[\protect\citeauthoryear{Fonseca, Camera, Santos  \& Maartens}{Fonseca
  et~al.}{2015}]{Fonseca:2015laa}
Fonseca J.,  Camera S.,  Santos M.,   Maartens R.,  2015, \mn@doi [Astrophys.
  J. Lett.] {10.1088/2041-8205/812/2/L22}, 812, L22

\bibitem[\protect\citeauthoryear{Giblin, Mertens  \& Starkman}{Giblin
  et~al.}{2016}]{Giblin:2016mjp}
Giblin J.~T.,  Mertens J.~B.,   Starkman G.~D.,  2016, \mn@doi [Astrophys. J.]
  {10.3847/1538-4357/833/2/247}, 833, 247

\bibitem[\protect\citeauthoryear{{G{\'o}rski}, {Hivon}, {Banday}, {Wandelt},
  {Hansen}, {Reinecke}  \& {Bartelmann}}{{G{\'o}rski}
  et~al.}{2005}]{2005ApJ...622..759G}
{G{\'o}rski} K.~M.,  {Hivon} E.,  {Banday} A.~J.,  {Wandelt} B.~D.,  {Hansen}
  F.~K.,  {Reinecke} M.,   {Bartelmann} M.,  2005, \mn@doi [\apj]
  {10.1086/427976}, \href {http://adsabs.harvard.edu/abs/2005ApJ...622..759G}
  {622, 759}

\bibitem[\protect\citeauthoryear{Green \& Wald}{Green \&
  Wald}{2012}]{Green:2011wc}
Green S.~R.,  Wald R.~M.,  2012, \mn@doi [Phys. Rev. D]
  {10.1103/PhysRevD.85.063512}, 85, 063512

\bibitem[\protect\citeauthoryear{Hahn, Angulo  \& Abel}{Hahn
  et~al.}{2015}]{Hahn_2015}
Hahn O.,  Angulo R.~E.,   Abel T.,  2015, \mn@doi [Mon. Not. Roy. Astron. Soc.]
  {10.1093/mnras/stv2179}, 454, 3920

\bibitem[\protect\citeauthoryear{Hassani, Adamek, Kunz  \& Vernizzi}{Hassani
  et~al.}{2019}]{Hassani:2019lmy}
Hassani F.,  Adamek J.,  Kunz M.,   Vernizzi F.,  2019, \mn@doi [JCAP]
  {10.1088/1475-7516/2019/12/011}, 12, 011

\bibitem[\protect\citeauthoryear{{Hockney} \& {Eastwood}}{{Hockney} \&
  {Eastwood}}{1981}]{1981csup.book.....H}
{Hockney} R.~W.,  {Eastwood} J.~W.,  1981, {Computer Simulation Using
  Particles}

\bibitem[\protect\citeauthoryear{Jelic-Cizmek, Lepori, Adamek  \&
  Durrer}{Jelic-Cizmek et~al.}{2018}]{Jelic_Cizmek_2018}
Jelic-Cizmek G.,  Lepori F.,  Adamek J.,   Durrer R.,  2018, \mn@doi [JCAP]
  {10.1088/1475-7516/2018/09/006}, 2018, 006–006

\bibitem[\protect\citeauthoryear{Jolicoeur, Umeh, Maartens  \&
  Clarkson}{Jolicoeur et~al.}{2018}]{Jolicoeur:2017eyi}
Jolicoeur S.,  Umeh O.,  Maartens R.,   Clarkson C.,  2018, \mn@doi [JCAP]
  {10.1088/1475-7516/2018/03/036}, 03, 036

\bibitem[\protect\citeauthoryear{Jolicoeur, Maartens, De~Weerd, Umeh, Clarkson
  \& Camera}{Jolicoeur et~al.}{2020}]{Jolicoeur:2020eup}
Jolicoeur S.,  Maartens R.,  De~Weerd E.~M.,  Umeh O.,  Clarkson C.,   Camera
  S.,  2020, preprint

\bibitem[\protect\citeauthoryear{Kaiser}{Kaiser}{1987}]{Kaiser:1987qv}
Kaiser N.,  1987, Mon. Not. Roy. Astron. Soc., 227, 1

\bibitem[\protect\citeauthoryear{Knabenhans et~al.}{Knabenhans
  et~al.}{2019}]{Knabenhans:2018cng}
Knabenhans M.,  et~al., 2019, \mn@doi [Mon. Not. Roy. Astron. Soc.]
  {10.1093/mnras/stz197}, 484, 5509

\bibitem[\protect\citeauthoryear{Laureijs et~al.}{Laureijs
  et~al.}{2011}]{Laureijs:2011gra}
Laureijs R.,  et~al., 2011, arXiv:1110.3193

\bibitem[\protect\citeauthoryear{Leclercq, Pisani  \& Wandelt}{Leclercq
  et~al.}{2014}]{Leclercq:2014jda}
Leclercq F.,  Pisani A.,   Wandelt B.~D.,  2014, \mn@doi [Proc. Int. Sch. Phys.
  Fermi] {10.3254/978-1-61499-476-3-189}, 186, 189

\bibitem[\protect\citeauthoryear{Leistedt, Peiris, Mortlock, Benoit-L\'evy  \&
  Pontzen}{Leistedt et~al.}{2013}]{Leistedt:2013gfa}
Leistedt B.,  Peiris H.~V.,  Mortlock D.~J.,  Benoit-L\'evy A.,   Pontzen A.,
  2013, \mn@doi [Mon. Not. Roy. Astron. Soc.] {10.1093/mnras/stt1359}, 435,
  1857

\bibitem[\protect\citeauthoryear{Lepori, Adamek, Durrer, Clarkson  \&
  Coates}{Lepori et~al.}{2020}]{Lepori:2020ifz}
Lepori F.,  Adamek J.,  Durrer R.,  Clarkson C.,   Coates L.,  2020, \mn@doi
  [Mon. Not. Roy. Astron. Soc.] {10.1093/mnras/staa2024}, 49, 2078–2095

\bibitem[\protect\citeauthoryear{Levi et~al.}{Levi et~al.}{2013}]{Levi:2013gra}
Levi M.,  et~al., 2013, arXiv:1308.0847

\bibitem[\protect\citeauthoryear{Lewis \& Challinor}{Lewis \&
  Challinor}{2006}]{Lewis:2006}
Lewis A.,  Challinor A.,  2006, \mn@doi [Physics Reports]
  {10.1016/j.physrep.2006.03.002}, 429, 1

\bibitem[\protect\citeauthoryear{Lorenz, Alonso  \& Ferreira}{Lorenz
  et~al.}{2018}]{Lorenz:2017iez}
Lorenz C.~S.,  Alonso D.,   Ferreira P.~G.,  2018, \mn@doi [Phys. Rev. D]
  {10.1103/PhysRevD.97.023537}, 97, 023537

\bibitem[\protect\citeauthoryear{Maartens, Jolicoeur, Umeh, De~Weerd, Clarkson
  \& Camera}{Maartens et~al.}{2020}]{Maartens:2019yhx}
Maartens R.,  Jolicoeur S.,  Umeh O.,  De~Weerd E.~M.,  Clarkson C.,   Camera
  S.,  2020, \mn@doi [JCAP] {10.1088/1475-7516/2020/03/065}, 03, 065

\bibitem[\protect\citeauthoryear{Macpherson, Lasky  \& Price}{Macpherson
  et~al.}{2017}]{Macpherson:2016ict}
Macpherson H.~J.,  Lasky P.~D.,   Price D.~J.,  2017, \mn@doi [Phys. Rev. D]
  {10.1103/PhysRevD.95.064028}, 95, 064028

\bibitem[\protect\citeauthoryear{Mead, Heymans, Lombriser, Peacock, Steele  \&
  Winther}{Mead et~al.}{2016}]{mead2016}
Mead A.~J.,  Heymans C.,  Lombriser L.,  Peacock J.~A.,  Steele O.~I.,
  Winther H.~A.,  2016, \mn@doi [Mon. Not. Roy. Astron. Soc.]
  {10.1093/mnras/stw681}, 459, 1468

\bibitem[\protect\citeauthoryear{Pueblas \& Scoccimarro}{Pueblas \&
  Scoccimarro}{2009}]{pueblas2009}
Pueblas S.,  Scoccimarro R.,  2009, \mn@doi [Phys. Rev. D]
  {10.1103/PhysRevD.80.043504}, 80, 043504

\bibitem[\protect\citeauthoryear{Raccanelli, Bertacca, Maartens, Clarkson  \&
  Dor\'e}{Raccanelli et~al.}{2016}]{Raccanelli:2013gja}
Raccanelli A.,  Bertacca D.,  Maartens R.,  Clarkson C.,   Dor\'e O.,  2016,
  \mn@doi [Gen. Rel. Grav.] {10.1007/s10714-016-2076-8}, 48, 84

\bibitem[\protect\citeauthoryear{Ramanah, Charnock, Villaescusa-Navarro  \&
  Wandelt}{Ramanah et~al.}{2020}]{Ramanah:2020vyl}
Ramanah D.~K.,  Charnock T.,  Villaescusa-Navarro F.,   Wandelt B.~D.,  2020,
  \mn@doi [Mon. Not. Roy. Astron. Soc.] {10.1093/mnras/staa1428}, 495, 4227

\bibitem[\protect\citeauthoryear{Rigopoulos \& Valkenburg}{Rigopoulos \&
  Valkenburg}{2015}]{Rigopoulos:2013nda}
Rigopoulos G.,  Valkenburg W.,  2015, \mn@doi [Mon. Not. Roy. Astron. Soc.]
  {10.1093/mnras/stu2070}, 446, 677

\bibitem[\protect\citeauthoryear{Spergel et~al.}{Spergel
  et~al.}{2015}]{Spergel:2015sza}
Spergel D.,  et~al., 2015, arXiv:1503.03757

\bibitem[\protect\citeauthoryear{Springel}{Springel}{2005}]{Springel:2005mi}
Springel V.,  2005, \mn@doi [Mon. Not. Roy. Astron. Soc.]
  {10.1111/j.1365-2966.2005.09655.x}, 364, 1105

\bibitem[\protect\citeauthoryear{Tansella, Bonvin, Durrer, Ghosh  \&
  Sellentin}{Tansella et~al.}{2018}]{Tansella:2017rpi}
Tansella V.,  Bonvin C.,  Durrer R.,  Ghosh B.,   Sellentin E.,  2018, \mn@doi
  [JCAP] {10.1088/1475-7516/2018/03/019}, 03, 019

\bibitem[\protect\citeauthoryear{Teyssier}{Teyssier}{2002}]{Teyssier:2001cp}
Teyssier R.,  2002, \mn@doi [Astron. Astrophys.] {10.1051/0004-6361:20011817},
  385, 337

\bibitem[\protect\citeauthoryear{Umeh, Jolicoeur, Maartens  \& Clarkson}{Umeh
  et~al.}{2017}]{Umeh:2016nuh}
Umeh O.,  Jolicoeur S.,  Maartens R.,   Clarkson C.,  2017, \mn@doi [JCAP]
  {10.1088/1475-7516/2017/03/034}, 03, 034

\bibitem[\protect\citeauthoryear{Umeh, Koyama, Maartens, Schmidt  \&
  Clarkson}{Umeh et~al.}{2019}]{Umeh:2019qyd}
Umeh O.,  Koyama K.,  Maartens R.,  Schmidt F.,   Clarkson C.,  2019, \mn@doi
  [JCAP] {10.1088/1475-7516/2019/05/020}, 05, 020

\bibitem[\protect\citeauthoryear{Wang, Beutler  \& Bacon}{Wang
  et~al.}{2020}]{Wang:2020ibf}
Wang M.~S.,  Beutler F.,   Bacon D.,  2020, \mn@doi [Mon. Not. Roy. Astron.
  Soc.] {10.1093/mnras/staa2998}, 499, 2598

\bibitem[\protect\citeauthoryear{Witzemann, Alonso, Fonseca  \&
  Santos}{Witzemann et~al.}{2019}]{Witzemann:2018cdx}
Witzemann A.,  Alonso D.,  Fonseca J.,   Santos M.~G.,  2019, \mn@doi [Mon.
  Not. Roy. Astron. Soc.] {10.1093/mnras/stz778}, 485, 5519

\bibitem[\protect\citeauthoryear{Yip et~al.,}{Yip et~al.}{2019}]{Yip:2019rfl}
Yip J.~H.,  et~al., 2019, in {33rd Annual Conference on Neural Information
  Processing Systems}.  (\mn@eprint {arXiv} {1910.07813})

\bibitem[\protect\citeauthoryear{Yoo \& Desjacques}{Yoo \&
  Desjacques}{2013}]{Yoo:2013tc}
Yoo J.,  Desjacques V.,  2013, \mn@doi [Phys. Rev. D]
  {10.1103/PhysRevD.88.023502}, 88, 023502

\bibitem[\protect\citeauthoryear{Yoo, Fitzpatrick  \& Zaldarriaga}{Yoo
  et~al.}{2009}]{Yoo:2009au}
Yoo J.,  Fitzpatrick A.,   Zaldarriaga M.,  2009, \mn@doi [Phys. Rev. D]
  {10.1103/PhysRevD.80.083514}, 80, 083514

\bibitem[\protect\citeauthoryear{Zonca, Singer, Lenz, Reinecke, Rosset, Hivon
  \& Gorski}{Zonca et~al.}{2019}]{Zonca2019}
Zonca A.,  Singer L.,  Lenz D.,  Reinecke M.,  Rosset C.,  Hivon E.,   Gorski
  K.,  2019, \mn@doi [Journal of Open Source Software] {10.21105/joss.01298},
  4, 1298

\makeatother
\end{thebibliography}

\end{document}